ARTICLE  OPEN 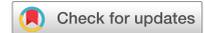

# Electronic stripe patterns near the fermi level of tetragonal Fe(Se,S)

M. Walker[1,2,3], K. Scott[2,3], T. J. Boyle[1,2,3], J. K. Byland[1], S. Bötzel[4], Z. Zhao[1], R. P. Day[5,6], S. Zhdanovich[5,6], S. Gorovikov[7], T. M. Pedersen[7], P. Klavins[1], A. Damascelli[5,6], I. M. Eremin[4], A. Gozar[2,3], V. Taufour[1] and E. H. da Silva Neto[1,2,3,8]✉

FeSe$_{1-x}$S$_x$ remains one of the most enigmatic systems of Fe-based superconductors. While much is known about the orthorhombic parent compound, FeSe, the tetragonal samples, FeSe$_{1-x}$S$_x$ with $x > 0.17$, remain relatively unexplored. Here, we provide an in-depth investigation of the electronic states of tetragonal FeSe$_{0.81}$S$_{0.19}$, using scanning tunneling microscopy and spectroscopy (STM/S) measurements, supported by angle-resolved photoemission spectroscopy (ARPES) and theoretical modeling. We analyze modulations of the local density of states (LDOS) near and away from Fe vacancy defects separately and identify quasiparticle interference (QPI) signals originating from multiple regions of the Brillouin zone, including the bands at the zone corners. We also observe that QPI signals coexist with a much stronger LDOS modulation for states near the Fermi level whose period is independent of energy. Our measurements further reveal that this strong pattern appears in the STS measurements as short range stripe patterns that are locally two-fold symmetric. Since these stripe patterns coexist with four-fold symmetric QPI around Fe-vacancies, the origin of their local two-fold symmetry must be distinct from that of nematic states in orthorhombic samples. We explore several aspects related to the stripes, such as the role of S and Fe-vacancy defects, and whether they can be explained by QPI. We consider the possibility that the observed stripe patterns may represent incipient charge order correlations, similar to those observed in the cuprates.

## INTRODUCTION

Iron-based superconductors (FeSCs) exhibit intertwined orders, including superconductivity, spin-density waves (SDWs), and nematicity. While most FeSC systems have neighboring magnetic and nematic phase transitions, the SDW phase is absent in FeSe$_{1-x}$S$_x$ at ambient pressure. The substitution of S for Se suppresses the tetragonal-to-orthorhombic transition temperature to zero at $x = x_c \approx 0.17$ (Fig. 1a), achieving a putative nematic quantum critical point (QCP)[1–3]. Unlike the prototypical 122 family of FeSCs, where the superconducting transition $T_c$ is maximum near the nematic/magnetic QCP, $T_c$ in Fe(Se, S) reaches a maximum at $x = 0.11$ and shows a suppression for $x > x_c$[3,4]. However, the superconductivity remains sensitive to the underlying orthorhombic or tetragonal structure, with many experiments showing an abrupt change in the superconducting properties across $x_c$[5,6]. The superconductivity at $x > x_c$ is not as well understood as in FeSe, leading to several hypotheses for exotic superconducting states, such as a Bose-Einstein condensate (BEC) phase and Bogoliubov Fermi surfaces, also called an ultranodal state[7–9]. Despite these differences across $x_c$, the normal state from which the superconductivity emerges is thought to be free of further electronic instabilities. In this context, we report scanning tunneling spectroscopy (STS) measurements, supported by angle-resolved photoemission spectroscopy (ARPES) and theoretical modeling, that probe the normal state of tetragonal Fe(Se, S), revealing that the local density of states (LDOS) forms strong stripe patterns for states just above the Fermi level.

Identifying broken symmetry states beyond superconductivity in tetragonal FeSe$_{1-x}$S$_x$ presents a challenge due to the random location of S. This is because the spatial inhomogeneity induced by S may limit the correlation length of such states, making their detection by spatially averaged experimental probes difficult. Moreover, the random S distribution may negatively impact spatially averaged probes of the band structure causing inhomogenous broadening effects on spectral features. This is observed, for example, in some ARPES measurements of tetragonal FeSe$_{1-x}$S$_x$, which yield broader band features than those observed in FeSe [3]. In light of these considerations, scanning tunneling microscopy and spectroscopy (STM/S) becomes a particularly suitable tool for investigating FeSe$_{1-x}$S$_x$ since it can probe electronic correlations and band structure with high spatial resolution.

We focus our studies on tetragonal FeSe$_{1-x}$S$_x$ samples with $x = 0.19$. We describe our various results in the following order: (i) STS imaging experiments over a $45 \times 45$ nm$^2$ area reveal distinct sets of quasiparticle interference (QPI) vectors when analyzing two distinct real-space regions. Specifically, areas near and away from the sparsely distributed Fe-vacancies (FeVs) yield different QPI features. Separating the Fourier transform analysis of these regions allows us to disentangle QPI features that are otherwise unresolved when the Fourier transform is done over the full field-of-view of our measurements. (ii) This scatterer-resolved QPI (SR-QPI) analysis enables a detailed comparison of the experimental data to theoretical Green's function calculations informed by ARPES measurements. This comparison allows us to identify the momentum space origin of various QPI scattering vectors. Notably, we identify QPI from the bands that form the electron pockets at the M/A points, which had not yet been identified in tetragonal Fe(Se, S). (iii) We find that the strongest periodic modulations observed in the STS measurements occur in the 0 to 10 meV energy range, along the nearest Fe-Fe distance, with an energy-independent wave vector ($\approx 0.12$ Å$^{-1}$) and in the regions away from FeVs. (iv) An analysis of this feature in real space reveals that it appears as alternating domains of short-range stripe patterns. Interestingly, we find that these stripe patterns coexist with four-fold symmetric QPI signals emanating from FeVs, whose locations are correlated with the boundaries between the stripe domains. Since QPI around FeVs are known to be two-fold symmetric within orthorhombic domains, the stripe patterns revealed in our measurements of tetragonal FeSe$_{1-x}$S$_x$ must be distinct from the nematic correlations in orthorhombic samples. Altogether, our observations suggest the presence of a periodic electronic modulation that breaks both translational and rotational symmetry in FeSe$_{0.81}$S$_{0.19}$.

[1]Department of Physics and Astronomy, University of California, Davis, CA, USA. [2]Department of Physics, Yale University, New Haven, CT, USA. [3]Energy Sciences Institute, Yale University, West Haven, CT, USA. [4]Institut für Theoretische Physik III, Ruhr-Universität Bochum, Bochum, Germany. [5]Quantum Matter Institute, University of British Columbia, Vancouver, BC, Canada. [6]Department of Physics & Astronomy, University of British Columbia, Vancouver, BC, Canada. [7]Canadian Light Source, Saskatoon, Saskatchewan, Canada. [8]Department of Applied Physics, Yale University, New Haven, CT, USA. ✉email: eduardo.dasilvaneto@yale.edu

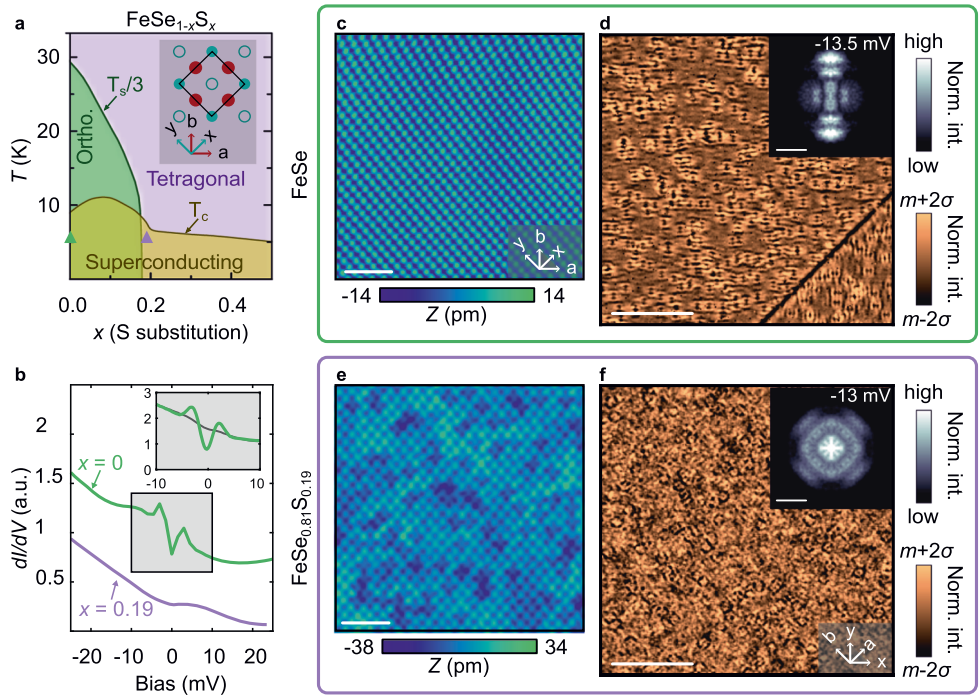

**Fig. 1 Overview of STS results comparing FeSe and FeSe$_{0.81}$S$_{0.19}$.** **a** Phase diagram of FeSe$_{1-x}$S$_x$ based on[3], indicating the structural and superconducting transitions, $T_S$ and $T_c$. For simplicity, we show the $T_S/3$ line. The two samples studied in this work ($x=0$ and $x=0.19$) were measured at 4.2 K, green and purple triangles, respectively. Inset is a schematic of the crystal structure of pure FeSe. Red circles are Fe, turquoise circles are Se one layer above, and turquoise open circles are Se in the layer below. **b** Representative spatially averaged differential conductance (d*I*/d*V*) spectra of FeSe (green) and FeSe$_{0.81}$S$_{0.19}$ (purple). The inset shows two spectra with higher energy resolution in the range of the superconducting gap, taken on the same FeSe sample with no field applied (green) and under 10 T (grey). **c**, **e** Representative constant current topographic images of FeSe and FeSe$_{0.81}$S$_{0.19}$ showing the atomically resolved (Se,S) termination layer. Scale bar represents 2 nm. **d**, **f** d*I*/d*V* maps on FeSe and FeSe$_{0.81}$S$_{0.19}$, at selected energies and over different fields-of-view than **c**, **e**. The scale bar represents 50 nm. The insets show the discrete Fourier transform (FT) of the real-space image. The scale bar inside the insets correspond to 0.1 Å$^{-1}$ and the **q**-space resolution is 0.004 Å$^{-1}$. The d*I*/d*V* maps were normalized where $m$ is the map mean and $\sigma$ is the standard deviation.

## RESULTS

Although our main findings relate to tetragonal FeSe$_{1-x}$S$_x$, we first compare STM/S measurements of FeSe and FeSe$_{0.81}$S$_{0.19}$ in Fig. 1, in order to describe the key differences between the two samples. Representative STM topographic images of the Se termination layers of FeSe and FeSe$_{0.81}$S$_{0.19}$ are shown in Fig. 1c, e. In FeSe$_{0.81}$S$_{0.19}$, the S atoms appear in the topography as densely distributed four-fold symmetric features that replace the Se atoms, consistent with previous experiments[5,10]. STS measurements provide additional information on the underlying electronic states and their symmetries. In FeSe, two-fold symmetric modulations of the density of states (DOS) oriented along the smallest Fe-Fe distances (a and b) are observed due to quasiparticle interference

(QPI) near underlying FeVs. These modulations reflect the C$_2$ symmetric band structure in the orthorhombic phase, as shown in Fig. 1d. The same images also reveal the presence of an orthorhombic domain wall, evidenced by the 90° rotated QPI patterns across the boundary, again consistent with previous measurements[11,12]. Fourier transforms (FTs) of the d*I*/d*V* patterns on the top left domain confirm their C$_2$ symmetry. In contrast, the FT patterns over a similar field of view in FeSe$_{0.81}$S$_{0.19}$ are C$_4$ symmetric, as shown in Fig. 1f. This observation is consistent with the sample being tetragonal on average.

### Distinct QPI near and away from Fe-vacancies

In Fig. 2 we present STS measurements over a range of energies on a 45 × 45 nm$^2$ field of view containing 8 FeV defects and, based on the size of the image and the percentage of S substitution, an estimated 2700 S atoms. At energies closer to the Fermi level (Fig. 2a, −15.25 meV) C$_4$ symmetric flower-like patterns are clearly seen centered around FeV defects on the top left of the image. At lower energies (Fig. 2b) the flower patterns shrink, showing the energy dependence that is a hallmark of QPI. Away from the FeV defects (bottom right region) periodic patterns are also observed, which, at −40 meV (Fig. 2c), pervade the entire image and dominate over the flower-like QPI signals. Comparing maps at −40 meV and −55 meV (Fig. 2c, d) one observes a decrease in the characteristic wavelength of these modulations, indicating they are also due to QPI. These observations motivate us to analyze regions around FeVs (orange shading in Fig. 2a) and vacancy poor areas (blue shading in Fig. 2a) separately. For certain energies, the FeVs appear as bright spots in the d*I*/d*V* map, which allows us to pinpoint their locations (see Supplementary Fig. 7a). Indeed,

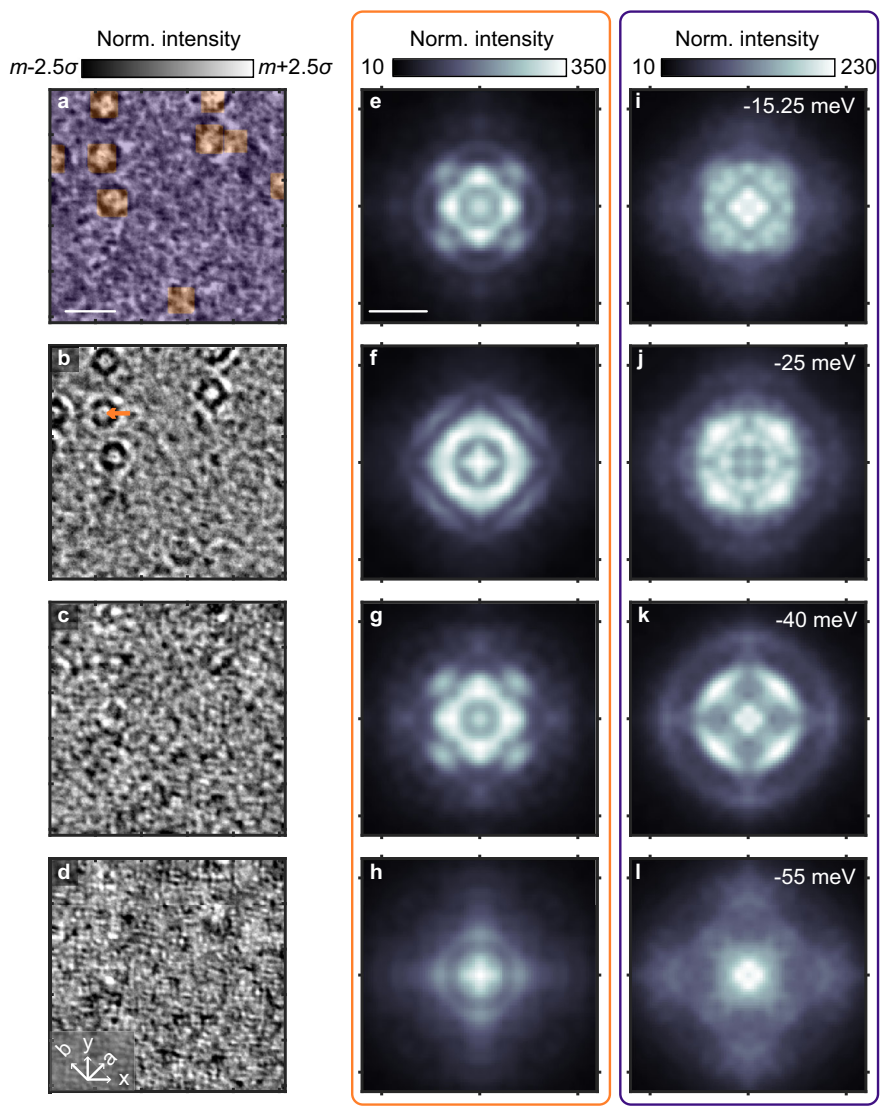

**Fig. 2 Distinct QPI near and away from FeVs. a–d** $dI/dV$ maps on FeSe$_{0.81}$S$_{0.19}$, at selected energies. The scale bar represents 10 nm. **e–h** FT of the near FeV region (orange) of the images in **a–d**. **i–l** FT of the off FeV region (blue) of the in **a–d**. The scale bar represents 0.3 Å$^{-1}$. The $dI/dV$ maps in **a–d** were normalized, where $m$ is the map mean and $\sigma$ is the standard deviation.

Fourier transforms over these distinct areas show significant differences, as can be seen for selected energies in Fig. 2e–l. Of course, a detailed understanding of the relative intensity between the QPI from the two regions requires knowledge of the spatial structure of their scattering, i.e. their form factors, but this is beyond the scope of this study. Still, the spatial masking analysis described above serves as a method to change the relative intensities of different QPI features.

The SR-QPI analysis introduced above has a clear practical utility in differentiating overlapping or nearby QPI features. This is particularly evident when analyzing dispersion maps, such as the ones shown in Figs. 3 and 4, which are obtained from the FTs of the STS maps, by taking line cuts along high-symmetry directions. In these dispersion maps the real-space $dI/dV$ images at each energy are normalized by their standard deviations before Fourier transformation. This normalization step helps track the momentum location of QPI features as a function of energy, even though it loses information about the relative intensity of features at different energies. Later we will present data without this energy-dependent normalization when comparing intensities.

To illustrate the practical result of the SR-QPI analysis, we show three different dispersion maps in Fig. 3a–c, generated over the full field-of-view (FOV), the orange region, and the blue region, as defined in Fig. 2a. Comparing the dispersion maps from the orange and blue regions, Fig. 3b, c, we observe clear differences in the −30 to −10 meV energy range. In the orange region, near FeVs, we observe a clear dispersive peak, marked by a black line, while in the blue region, away from FeVs, we observe two distinct dispersive peaks, marked by white lines. Superimposing those lines over the data shown in Fig. 3a, we can see that the three features merge into a single broad feature when analyzing the data over the full FOV. In the next section, we demonstrate how the extra resolution provided by the SR-QPI analysis significantly facilitates the identification of the scattering processes when comparing experiments to theory.

## Identifying the scattering processes in QPI

The wave vectors of QPI patterns at a given energy reflect the momentum transfer for scattering processes between states on the contours of constant energy of the band structure. To relate the STS data to the band structure, we use the low-energy model that describes the band structure in FeSe and FeSe$_{1-x}$S$_x$[13,14], fitted to our ARPES data for FeSe$_{0.77}$S$_{0.23}$, to simulate QPI patterns using a Green's function method (see Supplementary Equation 6). Figure 4a–d display the calculated dispersion maps. The scattering for the bands centered at the Γ and Z points are shown in Fig. 4a, b. Since the parameters used in our calculations result in circular pockets around Γ and Z in the $k_x$-$k_y$ plane, the simulations in Fig. 4a, b serve as predictions for both high-symmetry directions. The scattering for the bands centered at M/A points are shown in Fig. 4c, d. In this case, the QPI prediction is significantly different along the two high-symmetry directions (see Fig. 4c, d), but there is only negligible dispersion for those bands along $k_z$, meaning the calculations are nearly identical for M or A.

In principle, each experimental QPI feature can be matched to different lines in the theory simulations. However, in practice, this can be challenging because signals from different parts of the Brillouin zone overlap in the FTs of the STS data. We overcome this challenge with the SR-QPI analysis. Figure 4e–h show the experimental data along the two high-symmetry directions near (away) from FeVs. For each experimentally observed feature, we identified potential matches in the theory calculations, which are displayed as lines in Fig. 4e–h. In addition to scattering from the bands centered at Γ and Z (red and orange lines), some of which have already been identified by previous reports[5], our data also reveal experimental features that can only be attributed to intra-pocket scattering from the electron pockets at M/A points (represented by purple lines), which had not been resolved in QPI of FeSe$_{1-x}$S$_x$ before. These results not only demonstrate how STS measurements can access the electron pockets at the M/A points in FeSe$_{1-x}$S$_x$, but they also validate our use and interpretation of the SR-QPI analysis.

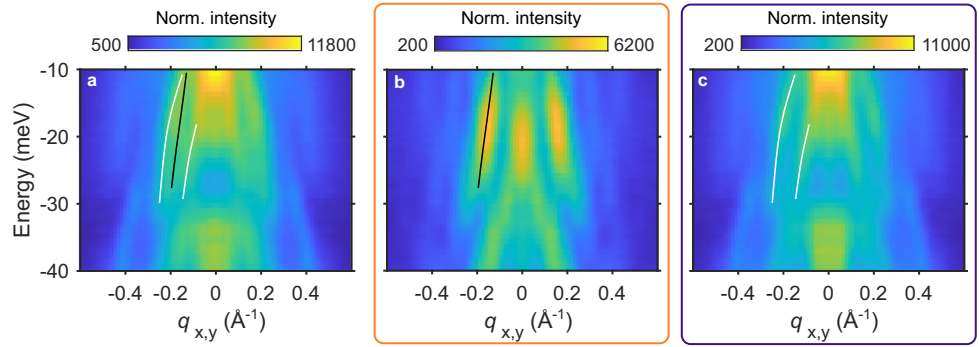

**Fig. 3 Illustration of the scatterer-resolved QPI analysis. a** Dispersion map along the Se-Se bond direction obtained from FTs taken over the full map in Fig. 2a. **b** Same as **a**, but over the orange region. **c** Same as **a**, but over the blue region. Lines drawn on **b** and **c** are guides to the eye highlighting bands in that region, and the same lines are shown in **a**.

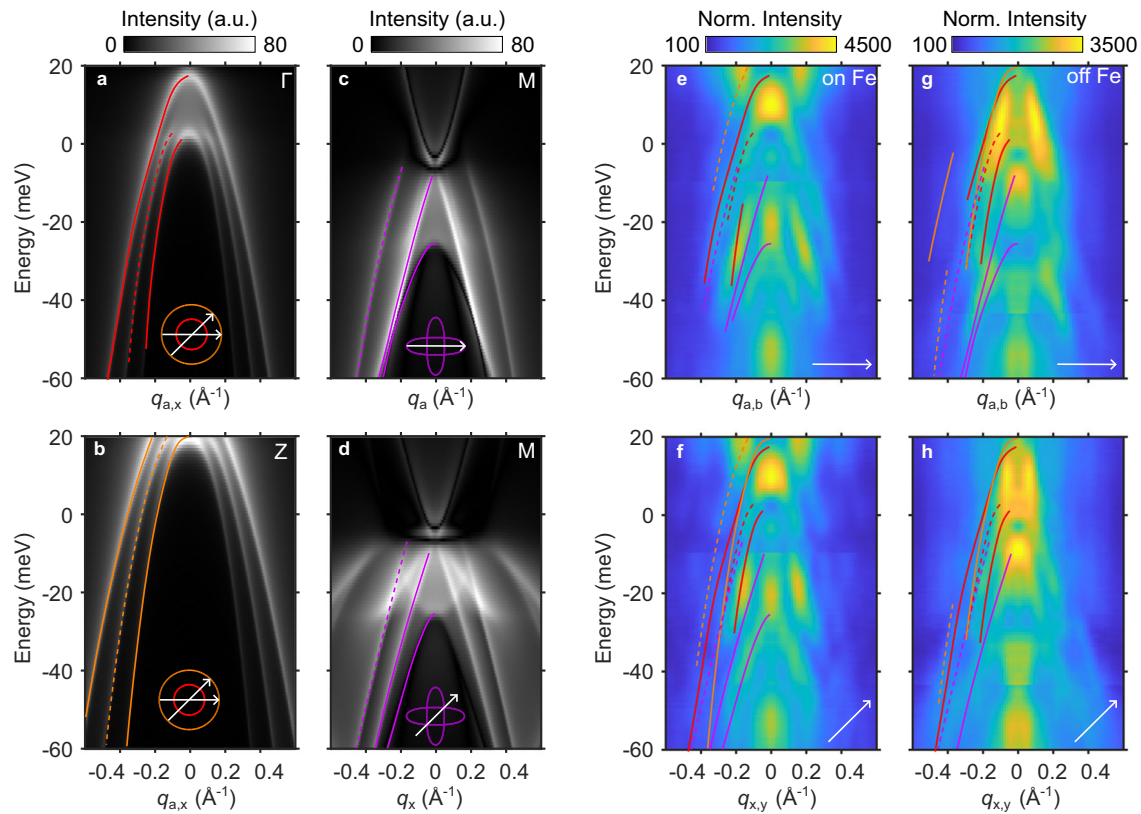

**Fig. 4 Identifying Scattering Processes. a–d** Calculated dispersion maps of QPI, along the directions indicated in the inset Brillouin zone schematic, obtained from a band structure model fitted to ARPES data. Red, orange, and purple lines are guides to the eye. A dashed line indicates that the feature is due to interband scattering. **e–h** Dispersion maps obtained from Fourier transforms over the orange region **e, f** and the blue region **g, h** of Fig. 2a along the directions indicated on the x-axis labels.

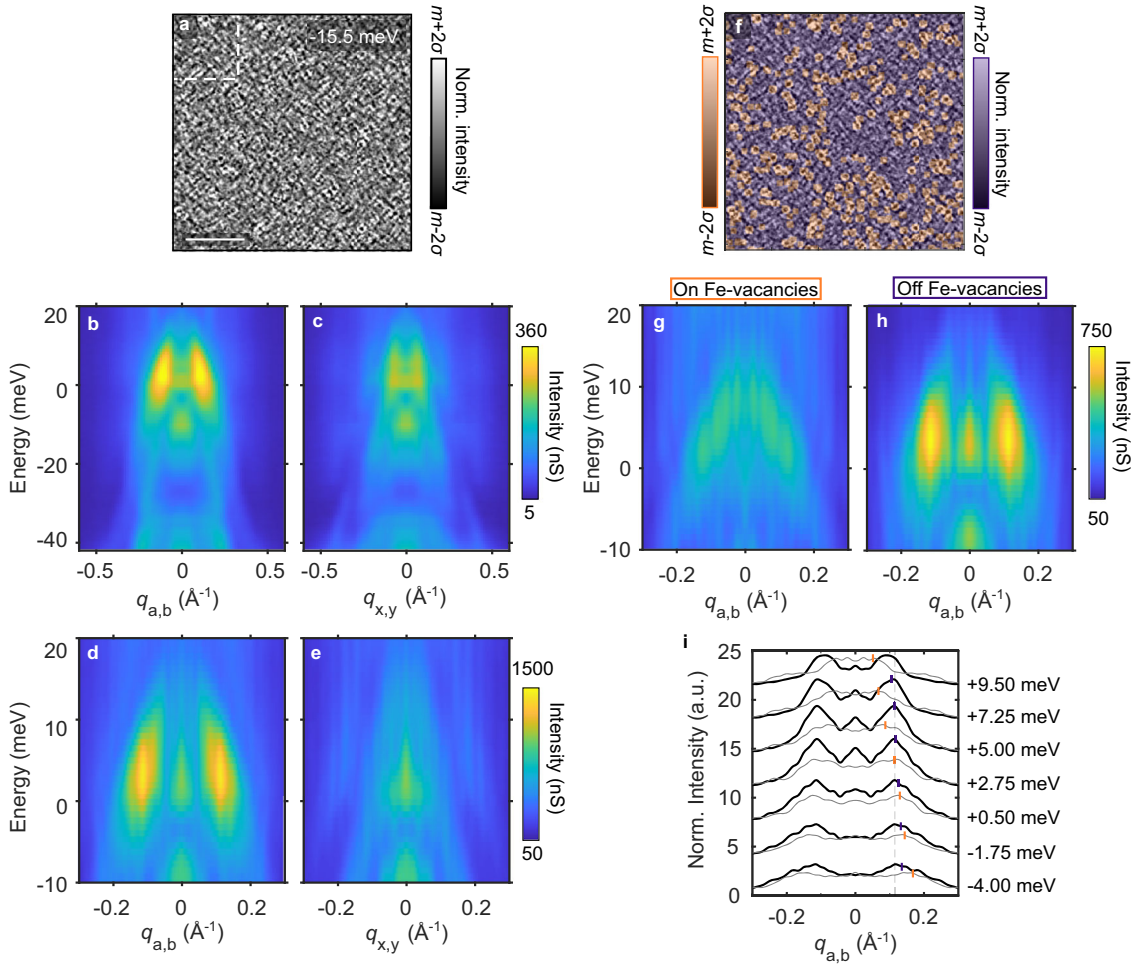

**Fig. 5 Identification of a non-dispersive signal above the Fermi level. a** $dI/dV$ map on FeSe$_{0.81}$S$_{0.19}$ at 6 meV. The scale bar represents 50 nm. **b**, **c** Dispersion maps along high-symmetry directions obtained from FTs of the 45 × 45 nm$^2$ FOV (see Fig. 2). **d**, **e** Similar to **b**, **c** but obtained from the larger 170 × 170 nm$^2$ FOV shown in **a**, and with higher energy resolution. **f** Same as **a** but with transparent masks separating the map into two regions: near FeVs (orange), and away from FeVs (blue). **g**, **h** Dispersion maps along the smallest Fe-Fe direction, obtained from FTs over the orange and blue regions in **f**, respectively. These two images were scaled so their intensities sum to that of the full map. **i** Constant energy line cuts of the dispersion maps in **h** (gray lines) and **i** (black lines). The orange markers are the locations of the peaks obtained from fits to the linecuts (see Supplementary Fig. 5). The FTs used to make the dispersion maps in this figure were performed over the mean subtracted $dI/dV$ maps without normalization.

### Non-dispersive broken-symmetry signal above the Fermi level

Despite the success of our theoretical modeling in capturing most of the experimental QPI features, the calculations are unable to explain the signal just above the Fermi level. The feature between 0 and 10 meV, near $q_{a,b} = 0.12$ Å$^{-1}$, stands out as the most intense signal in our STS measurements, which is evident in the dispersion maps obtained from the entire FOV of Fig. 2 (dashed box in Fig. 5a) but without using the energy-dependent normalization (see Fig. 5b, c). The prominence of this feature suggests that there might be more at play than simply QPI, which prompted us to further investigate it with higher energy and momentum resolution (Fig. 5d, e). We achieved the latter by making STS measurements over larger FOVs (170 × 170 nm$^2$, see Fig. 5a). Fourier transforming over the larger FOV suggests that the ≈ 0–10 meV region hosts not only a high intensity feature with $q_{a,b} = 0.12$ Å$^{-1}$ but also a weaker feature underneath with hole-like dispersion. This could either be a single dispersive feature whose intensity is enhanced around ≈ 5 meV, or two distinct features. Application of the SR-QPI analysis (see Fig. 5f) clarifies the situation, revealing two overlapping features, one dispersive (Fig. 5g), originating from FeVs, and one non-dispersive (Fig. 5h), originating from regions without FeVs. The same conclusions can be reached by examining individual line cuts of these data for various energies (see Fig. 5i). The coexistence of dispersive and non-dispersive features are typical in systems with charge density waves (CDWs), where the non-dispersive excitation reflects CDW correlations[15,16]. Therefore, with the same phenomenology being observed in our data, our results suggest the presence of electron correlations that break translational symmetry in FeSe$_{0.81}$S$_{0.19}$.

To better understand the non-dispersive modulations, we can examine their local rotational symmetry. Figure 6a shows the large FOV STS map at 6 meV, where the non-dispersive excitation is most prominent, revealing modulations that span the entire map. Figure 6b shows a zoom-in of the region delimited by the white box in Fig. 6a, after filtering the image to highlight the local structure of the 0.12 Å$^{-1}$ modulations. Here one can already see that locally the modulations appear as stripe patterns that alternate direction by 90° degrees. This is confirmed in Fig. 6c, which shows that the large FOV map in Fig. 6a can be separated into two regions (red and blue) whose individual FTs show 0.12 Å$^{-1}$ modulations only along $q_a$ or only along $q_b$.

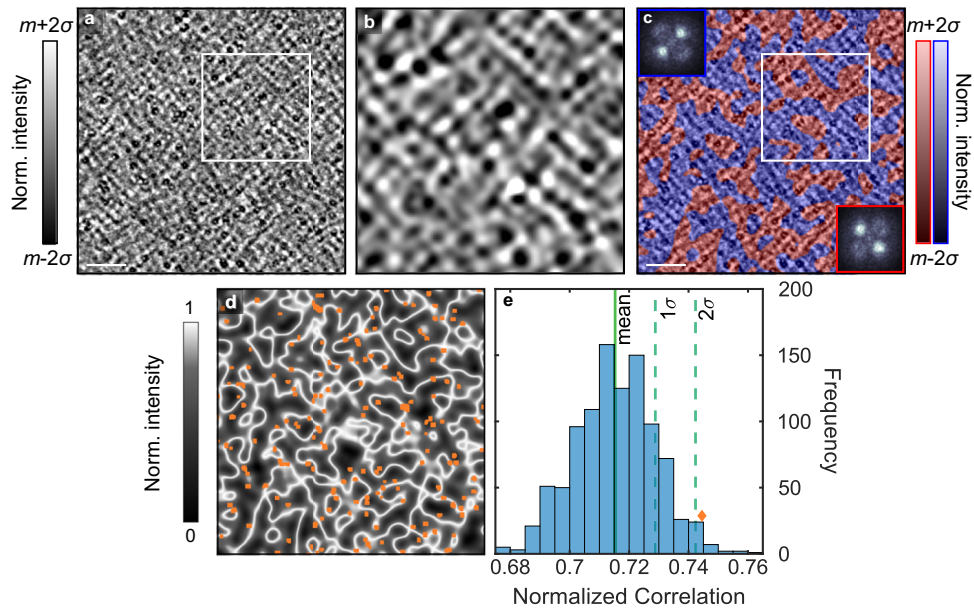

**Fig. 6 Local rotational symmetry analysis of the non-dispersive q = 0.12 Å$^{-1}$ feature. a** $dI/dV$ map measured on a FeSe$_{0.81}$S$_{0.19}$ sample at 6 mV. The scale bar represents 25 nm. **b** Magnified view of the region marked by the square in **a**. The image is filtered to highlight the local structure of the 0.12 Å$^{-1}$ modulations (See Supplementary Fig. 6). **c** The image in **a** overlaid with red and blue transparent masks that indicate regions of 0.12 Å$^{-1}$ modulations only along $a$ or only along $b$. The insets show the FTs over the two distinct regions under the red and blue masks. **d** Map indicating the domain boundaries between the red and blue regions (white) with locations of iron-vacancies overlaid (orange). **e** Histogram of the correlation between the location of FeVs and the domain boundaries in **d**, obtained from 1000 simulated sets of 150 randomly located FeVs. The orange diamond indicates the correlation between the experimental iron-vacancy locations and the domain boundaries. The solid green line represents the mean of the distribution and the two dashed lines represent 1 and 2 standard deviations, $\sigma$.

Next we examine the role of FeVs in the formation of the stripes patterns shown in Fig. 6c by looking at their location relative to the domain boundaries. We note that, based on the area of the image and the number of FeVs present, the average distance between FeVs is ≈ 150 Å, larger than the wavelength of the non-dispersive modulations ($\lambda \approx 52$ Å). Figure 6d overlays the location of FeVs on a map of the domain boundaries. To evaluate the connection between the FeVs and the domain walls, we calculate a correlation value between the two and compare that value to a distribution of correlation values obtained from 1000 motifs of randomly located FeVs (see Supplementary Fig. 7). This comparison is displayed in Fig. 6 and shows that the cross-correlation from experiments is approximately two standard deviations higher than the average cross-correlation obtained from randomly distributed sets of vacancies. Our analysis shows that FeVs are concentrated near the domain boundaries, therefore serving as local pinning sites.

## DISCUSSION

Our theoretical modeling informed by ARPES measurements is successful at explaining in detail the observed QPI patterns, despite the intricate band dispersions inherent to this multi-band system. However, there remains an outstanding feature that calculations could not capture: the non-dispersive excitation shown in Fig. 5 which also happens to be the feature with the strongest intensity in our data. In the following, we focus our discussion on the possible origins of this excitation.

Stripe-like modulations in FeSCs with different periods and coherence lengths have been reported by STM/S in two other situations: LiFeAs under external uniaxial stress[17] and ultrathin (one and two monolayers) FeSe films on SrTiO$_3$[18,19]. However, in those two cases, the underlying crystal structure already breaks rotational symmetry, indicating that a C$_2$ symmetric lattice may be necessary to foster the formation of those stripes. Instead, we find C$_2$ symmetric stripe patterns that are present in the tetragonal samples, FeSe$_{0.81}$S$_{0.19}$, but not in the orthorhombic case, FeSe.

Since nematic fluctuations persist with S-substitution in tetragonal FeSe$_{1-x}$S$_x$ samples[2,20,21], a natural hypothesis is that the observed stripe domains in FeSe$_{0.81}$S$_{0.19}$ represent a vestigial short-ranged order inherited from the parent system. One can imagine that the addition of S destroys the long-ranged nematic order and although the samples are tetragonal on average, local uniaxial strain fields stabilize the formation of short-range nematic domains. However, the evolution of the QPI patterns around FeVs from FeSe to FeSe$_{0.81}$S$_{0.19}$ (see Figs. 1 and 2) suggest a different scenario. Specifically, because the C$_2$ symmetry of the underlying lattice is strongly manifested in the QPI patterns around FeVs in FeSe crystals, we would expect that to also remain true in the S-substituted systems. This is opposite to what we see: the QPI in FeSe$_{0.81}$S$_{0.19}$ are C$_4$ symmetric while the observed stripe patterns are stronger in regions where Fe stoichiometry is locally preserved. This argument indicates that the mechanism behind the rotational symmetry breaking of the stripe patterns must be distinct from that of the nematic states in orthorhombic FeSe.

As the non-dispersive feature is not present in S-free crystals it is also important to assess the possible relevance of disorder to the characteristic length scale of the stripe modulations. The wavelength of these modulations, which is about 52 Å, is much larger than the average distance between S atoms, which is ≈ 8 Å. The discrepancy between these numbers, combined with the fact that S atoms sit at C$_4$ symmetric locations on the lattice, suggests that, at the length scale of the non-dispersive modulation wavelength, the random S may simply be thought of as a weak and quasi-isotropic strain field. As far as Fe vacancies are concerned, it is reasonable to consider whether the local two-fold symmetric patterns could arise from a random configuration of these vacancies. In principle, a possible explanation is that four-fold symmetric QPI patterns emanating from adjacent FeVs could interfere in a way that results in two-fold symmetric patterns in the intermediate regions. However, in this simple scenario, one would expect the modulations to be strongest near the FeVs and to progressively dampen out the further one moves away from them. This contradicts a salient aspect of our data, where the intensity of the non-dispersive modulations is strongest in the regions between the FeVs.

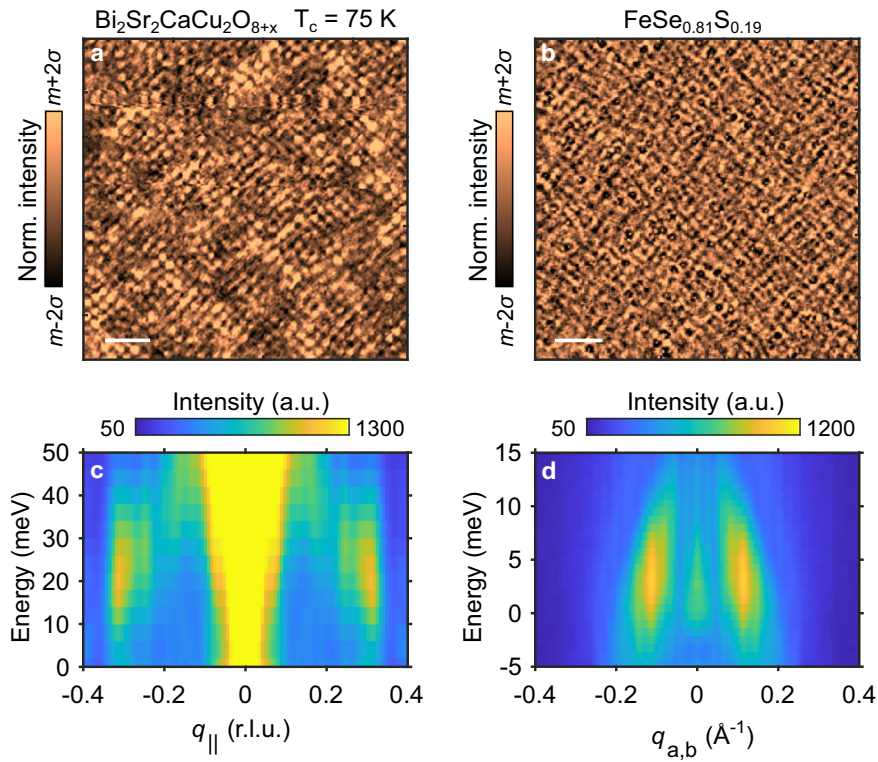

**Fig. 7 Comparison of charge order in Bi$_2$Sr$_2$CaCu$_2$O$_{8+x}$. a** d$I$/d$V$ map of a cuprate superconductor Bi$_2$Sr$_2$CaCu$_2$O$_{8+x}$ with $T_c$ = 75 K at +20 meV. The scale bar represents 5 nm. **b** d$I$/d$V$ map at +6 meV on FeSe$_{0.81}$S$_{0.19}$. The scale bar represents 25 nm. **c** Dispersion map obtained from the FTs of d$I$/d$V$ measurements on Bi$_2$Sr$_2$CaCu$_2$O$_{8+x}$ with $q_\parallel$ along the Cu-O bond direction. **d** Dispersion map obtained from the FTs of d$I$/d$V$ measurements on FeSe$_{0.81}$S$_{0.19}$ along the smallest Fe-Fe direction. Solid and dashed black lines are guides to the eye highlighting the dispersive and non-dispersive features, respectively. For more details on the data in **a** and **c**, see Ref. [16] where they were first published.

The phenomenology of the strong, non-dispersive feature is, however, remarkably similar to that of broken-symmetry states in cuprate superconductors where the existence of a stripe-like charge ordering (CO) state is well established. For example, the CO in Bi$_2$Sr$_2$CaCu$_2$O$_{8+x}$ appears in the real-space d$I$/d$V$ as local domains of alternating stripes[22], similar to the broken rotational and translational symmetry manifest in the d$I$/d$V$ maps of FeSe$_{0.81}$S$_{0.19}$. Furthermore, the energy momentum structure shows that the the signature of the CO in Bi$_2$Sr$_2$CaCu$_2$O$_{8+x}$ is a non-dispersive wavevector over an energy range just above the Fermi level, where it coexists with a much less intense dispersive QPI feature[16]. Figure 7 shows a side-by-side comparison of the charge order in Bi$_2$Sr$_2$CaCu$_2$O$_{8+x}$ and FeSe$_{0.81}$S$_{0.19}$. This comparison demonstrates that, while the length and energy scales are different, the same three phenomenological features appear in both materials: (i) the real-space d$I$/d$V$ appears as an alternating pattern of uni-directional stripes (ii) the energy momentum structure has an intense feature at finite **q** which does not disperse with energy (dashed black lines), and (iii) this non-dispersive feature appears at an energy window and wavevector near dispersive QPI features (solid black lines). Altogether, this striking similarity to cuprates strongly suggests that the origin of the non-dispersive feature in our data is due to the presence of local charge order in FeSe$_{1-x}$S$_x$. An interesting characteristic of the non-dispersive CO feature in the cuprates is its energy-dependent intensity, similar to our observations in FeSe$_{0.81}$S$_{0.19}$. In the context of the cuprates, theoretical works show that such an energy-dependent intensity is expected when considering the effects of impurity-pinned dynamic density-wave correlations on the STS measured LDOS, and that an energy enhancement may occur when underlying QPI features overlap with the ordering wavevector[23–25]. Therefore, the energy-dependent intensity of the non-dispersive feature in FeSe$_{0.81}$S$_{0.19}$ may indicate the observation of dynamic CO correlations pinned by impurities. An additional observation from our studies is that FeVs defects appear near domain walls of the stripe modulations, as explained above, possibly further limiting the coherence length of the CO. Extending the above phenomenological comparison one can speculate that the stripe modulations in FeSe$_{1-x}$S$_x$ could also emerge from strong electron correlations, which are present in Fe-based superconductors in an orbital selective manner[26–32].

Existing theoretical work provides further insight into why the FeSe$_{1-x}$S$_x$ system could be unstable towards a CO state. On one hand existing calculations show that electronic compressibility in FeSe crystals is enhanced by decreased electronic correlations[33]. At the same time, band renormalizations, as inferred from ARPES data[3,34], indicate that correlations are getting suppressed with increasing S concentration. Therefore, isoelectronic S substitution for Se could push the system into a region of diverging compressibility, leading to instability with respect to either phase separation or incommensurate CO. From a different theoretical perspective, an emerging ground state with unidirectional charge modulations could be related to the generic presence of anisotropic defects at the surface, such as atomic terrace step-edges[35]. In this picture, the nematic order parameter coupled to a weakly-screened uniaxial strain field, induced by the surface disorder, results in the stabilization of a smectic phase. This phase would be more stable than the bulk nematic order and would appear at a small but finite characteristic wavevector, depending on the nematoelastic coupling and surface roughness. The large wavelength predicted by this theory may explain the recent observations of modulated nematic order in FeSe and BaFe$_2$(As$_{0.87}$P$_{0.13}$)$_2$ by photoemission electron microscopy[36]. However, one can also imagine that the potentially large period over which the nematic phase is modulated in real-space may be reduced by the presence of atomic defects, such as FeVs.

In conclusion, our STS experiments and SR-QPI analysis represent significant progress in the measurements of QPI features, band structure mapping and detection of broken symmetry states in FeSe$_{1-x}$S$_x$ samples. Informed by ARPES, our QPI simulations are successful at explaining almost all features observed in our STS dispersion maps, but fail to capture the strongest STS signal, which is present along the shortest Fe-Fe crystallographic direction at $q_{a,b} \approx 0.12$ Å$^{-1}$ and occur for energies just above the Fermi level. Our STS measurements further reveal that these modulations appear in the form of short-range stripe patterns, whose origin is unrelated to the nematic order in orthorhombic FeSe$_{1-x}$S$_x$. Phenomenological similarities between STS observations in FeSe$_{0.81}$S$_{0.19}$ and in Bi$_2$Sr$_2$CaCu$_2$O$_{8+x}$ suggest that the stripe modulations reported here may be due to incipient charge order correlations, a scenario that finds support in existing theoretical work.

## METHODS

### Sample growth and characterization

Single crystals of FeSe and FeSe$_{1-x}$S$_x$ were grown using the chemical vapor transport method in a tilted furnace following the methods outlined in ref. [37]. Pure samples were characterized using a powder x-ray diffractometer. To determine the relationship between actual and nominal sulfur substitution levels, samples were characterized using a scanning electron microscope equipped with an energy dispersive x-ray (EDX) spectroscopy probe and counting S atoms in STM topographies. Several platelets with dimensions between 1–2 mm and 1-2 mm were selected from each batch for characterization.

### STM/S measurements

STM/S measurements were done with a customized Unisoku USM-1300 instrument. The samples were cleaved in situ in an ultra-high vacuum environment with pressures below $10^{-9}$ Torr. All STM/S measurements were performed at 4.2 K. Differential conductance measurements ($dI/dV$) were performed using a lock-in technique. Tunneling current and bias setpoint conditions, as well as lock-in parameters, are summarized in Supplementary Table 1. All Fourier transformations shown in this work were performed using a discrete fast Fourier transform algorithm on a normalized $dI/dV$ map, unless stated otherwise in the figure caption. The equation for the Fourier transform used is given by $Y_{p+1,q+1} = \sum_{j=0}^{m-1} \sum_{k=0}^{n-1} e^{-2\pi i \frac{jp}{m}} e^{-2\pi i \frac{kq}{n}} X_{j+1,k+1}$, where j, k and p, q label the pixels of the original and resultant images, respectively. Details for ARPES measurements and QPI simulations can be found in the Supplementary Methods.

## DATA AVAILABILITY

The data that support the findings of this study are available from the corresponding authors upon reasonable request.

## ACKNOWLEDGEMENTS
This material is based upon work supported by the National Science Foundation under Grants Nos. 1845994 and 2034345. The synthesis was supported by the UC Laboratory Fees Research Program (LFR-20-653926). Some of the research described in this paper was carried out at the Canadian Light Source, a national research facility of the University of Saskatchewan, which is supported by the Canada Foundation for Innovation (CFI), the Natural Sciences and Engineering Research Council (NSERC), the National Research Council (NRC), the Canadian Institutes of Health Research (CIHR), the Government of Saskatchewan, and the University of Saskatchewan. This research was undertaken thanks in part to funding from the Max Planck-UBC-UTokyo Centre for Quantum Materials and the Canada First Research Excellence Fund, Quantum Materials and Future Technologies Program. This project is also funded by the Canada Research Chairs Program (A.D.); the British Columbia Knowledge Development Fund (BCKDF); and the CIFAR Quantum Materials Program. This work was supported by the Alfred P. Sloan Fellowship (E.H.d.S.N.). We thank Matteo Michiardi for technical guidance and support regarding the ARPES measurements.

## AUTHOR CONTRIBUTIONS
M.W. and E.H.D.S.N. designed the experiment. M.W., K.S., T.J.B., and E.H.D.S.N. performed the STM measurements with the assistance of Z.Z., A.G., and P.K. M.W. and S.Z. performed the ARPES measurements with the assistance of R.P.D., S.G., and T.M.P. and the support of A.D. S.B., I.M.E., and A.G. performed the LDOS calculations with assistance from M.W. and E.H.D.S.N. J.K.B. grew and characterized the $FeSe_{1-x}S_x$ crystals with support from V.T. M.W. performed the data analysis in discussion with E.H.D.S.N. and A.G. M.W., E.H.D.S.N., and A.G. wrote the manuscript with input from all other authors.

## COMPETING INTERESTS
The authors declare no competing interests.

## ADDITIONAL INFORMATION
**Supplementary information** The online version contains supplementary material available at https://doi.org/10.1038/s41535-023-00592-5.

**Correspondence** and requests for materials should be addressed to E. H. da Silva Neto.

**Reprints and permission information** is available at http://www.nature.com/reprints

**Publisher's note** Springer Nature remains neutral with regard to jurisdictional claims in published maps and institutional affiliations.

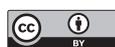